\documentclass[prx,twocolumn,showpacs,floatfix,superscriptaddress,amsmath,amsfonts,amssymb,preprintnumbers,citeautoscript,aps,longbibliography]{revtex4-1}
\pdfoutput=1
\usepackage[T1]{fontenc}\usepackage[latin1]{inputenc}
\usepackage{dcolumn,graphicx,color,booktabs,microtype,afterpage} \graphicspath{{fig/}}
\usepackage[charter,greekuppercase=italicized]{mathdesign}\usepackage{sidecap}
\usepackage[colorlinks,plainpages=false,linkcolor=blue,urlcolor=blue,citecolor=blue,pdfpagemode=UseNone,pdfstartview=FitBH]{hyperref}

\begin{document}
\title{Frustration model and spin excitations in the helimagnet FeP}
\author{A.~S.~Sukhanov}
\affiliation{Institut f{\"u}r Festk{\"o}rper- und Materialphysik, Technische Universit{\"a}t Dresden, D-01069 Dresden, Germany}
\author{Y.~V.~Tymoshenko}\altaffiliation[Now at ]{Institute for Quantum Materials and Technologies, Karlsruhe Institute of Technology, 76131 Karlsruhe, Germany}
\affiliation{Institut f{\"u}r Festk{\"o}rper- und Materialphysik, Technische Universit{\"a}t Dresden, D-01069 Dresden, Germany}
\author{A.~A.~Kulbakov}
\affiliation{Institut f{\"u}r Festk{\"o}rper- und Materialphysik, Technische Universit{\"a}t Dresden, D-01069 Dresden, Germany}
\author{A.~S.~Cameron}
\affiliation{Institut f{\"u}r Festk{\"o}rper- und Materialphysik, Technische Universit{\"a}t Dresden, D-01069 Dresden, Germany}
\author{V.~Kocsis}
\affiliation{Institut f\"{u}r Festk\"{o}rperforschung, Leibniz IFW-Dresden, D-01069 Dresden, Germany}
\author{H.~C.~Walker}
\affiliation{ISIS Facility, STFC, Rutherford Appleton Laboratory,
Didcot, Oxfordshire OX11-0QX, United Kingdom}
\author{A.~Ivanov}
\affiliation{Institut Laue-Langevin, 71 avenue des Martyrs, CS 20156, 38042 Grenoble Cedex 9, France}
\author{J.~T.~Park}
\affiliation{Heinz Maier-Leibnitz Zentrum (MLZ), TU M{\"u}nchen, D-85747 Garching, Germany}
\author{V. Pomjakushin}
\affiliation{Laboratory for Neutron Scattering and Imaging (LNS), Paul Scherrer Institute (PSI), CH-5232 Villigen, Switzerland}
\author{S. E. Nikitin}
\affiliation{Paul Scherrer Institute (PSI), CH-5232 Villigen, Switzerland}
\author{I.~V.~Morozov}
\affiliation{Institut f\"{u}r Festk\"{o}rperforschung, Leibniz IFW-Dresden, D-01069 Dresden, Germany}
\author{I.~O.~Chernyavskii}
\affiliation{Institut f\"{u}r Festk\"{o}rperforschung, Leibniz IFW-Dresden, D-01069 Dresden, Germany}
\author{S.~Aswartham}
\affiliation{Institut f\"{u}r Festk\"{o}rperforschung, Leibniz IFW-Dresden, D-01069 Dresden, Germany}
\author{A.~U.~B.~Wolter}
\affiliation{Institut f\"{u}r Festk\"{o}rperforschung, Leibniz IFW-Dresden, D-01069 Dresden, Germany}
\author{A.~Yaresko}
\affiliation{Max-Planck-Institut f\"{u}r Festk\"{o}rperforschung, Heisenbergstrasse 1, D-70569 Stuttgart, Germany}
\author{B.~B\"uchner}
\affiliation{Institut f{\"u}r Festk{\"o}rper- und Materialphysik, Technische Universit{\"a}t Dresden, D-01069 Dresden, Germany}
\affiliation{Institut f\"{u}r Festk\"{o}rperforschung, Leibniz IFW-Dresden, D-01069 Dresden, Germany}
\affiliation{W\"urzburg-Dresden Cluster of Excellence on Complexity and Topology in Quantum Matter\,---\,\textit{ct.qmat}, Technische Universit{\"a}t Dresden, 01069 Dresden, Germany}
\author{D.~S.~Inosov}\thanks{corresponding author: dmytro.inosov@tu-dresden.de}
\affiliation{Institut f{\"u}r Festk{\"o}rper- und Materialphysik, Technische Universit{\"a}t Dresden, D-01069 Dresden, Germany}
\affiliation{W\"urzburg-Dresden Cluster of Excellence on Complexity and Topology in Quantum Matter\,---\,\textit{ct.qmat}, Technische Universit{\"a}t Dresden, 01069 Dresden, Germany}
\begin{abstract}

The metallic compound FeP belongs to the class of materials that feature a complex noncollinear spin order driven by magnetic frustration. While its double-helix magnetic structure with a period $\lambda_{\text{s}} \approx 5c$, where $c$ is the lattice constant, was previously well determined, the relevant spin-spin interactions that lead to that ground state remain unknown.  By performing extensive inelastic neutron scattering measurements, we obtained the spin-excitation spectra in a large part of the momentum-energy space. The spectra show that the magnons are gapped with a gap energy of $\sim$5~meV. Despite the 3D crystal structure, the magnon modes display strongly anisotropic dispersions, revealing a quasi-one-dimensional character of the magnetic interactions in FeP. The physics of the material, however, is not determined by the dominating exchange, which is ferromagnetic. Instead, the weaker two-dimensional antiferromagnetic interactions between the rigid ferromagnetic spin chains drive the magnetic frustration. Using linear spin-wave theory, we were able to construct an effective Heisenberg Hamiltonian with an anisotropy term capable of reproducing the observed spectra. This enabled us to quantify the exchange interactions in FeP and determine the mechanism of its magnetic frustration.

\end{abstract}

\maketitle

\section{Introduction}
\subsection{General motivation}

The orthorhombic compound FeP (space group $Pnma$) belongs to the group of frustration-driven helimagnets. Unlike the noncentrosymmetric materials that exhibit a chiral (either left- or right-handed) spin-spiral magnetic order of the same handedness in the entire volume due to the antisymmetric spin-orbit-coupling dependent Dzyaloshinskii-Moriya interaction (DMI)~\cite{Muehlbauer09,Moskvin13,Adams12,Tokunaga15,Bogdanov94}, the frustration-driven helimagnets retain the degeneracy between the left- and right-handed spin spirals~\cite{Singh16,Jiang20,Zajdel17,Kim14,Inosov20}. In other words, competing exchange interactions between neighboring spins in the crystal lattice of a material set the period of the spin helix but not its handedness. The latter leads to nucleation of domains with the opposite chirality in a macroscopic sample. The fundamental difference in the underlying mechanisms of the formation of spin spirals in chiral and achiral materials manifests itself in their excitation spectrum.

Whilst the excitations of the chiral (DMI-based) helimagnets were broadly discussed in recent studies~\cite{Kugler15,Weber18,Che21,Aqeel21}, the magnon spectrum of the centrosymmetric (achiral) spin-spiral materials remains less studied~\cite{Rule17,Masuda05,Tymoshenko17}. Chiral helimagnets usually feature long spiral periods (much greater than the unit cell parameter), which is dictated by the smallness of the DMI as compared to the isotropic Heisenberg exchange. This allows for an effective continuous theory to be applied, which yields generalized results applicable for a wide range of real materials~\cite{Schwarze15,Garst17}. In contrast, the frustrated exchange interactions responsible for the spin-spiral ground state are typically of the same magnitude and lead to relatively short spiral pitches (of only a few unit cells). This requires that case-specific microscopic models be constructed for a particular material.

In this paper, we study the helimagnon spectrum of FeP by means of inelastic neutron scattering (INS) and propose a relevant spin-spin interaction model that explains the ground-state magnetic structure of the compound as well as its spin excitation spectrum. The paper is organized as follows. In Sec.~\ref{magn_struc}, we briefly overview the previously reported magnetic structure of FeP. In Sec.~\ref{frust_trapez}, we review and discuss the spin model that was previously proposed as the candidate model for FeP. Section~\ref{exp_det} summarizes the results of sample characterization and the experimental details of the performed neutron measurements. The results of the time-of-flight (TOF) INS measurements are presented in Sec.~\ref{TOF} along with the comparison to the model of Sec.~\ref{frust_trapez}. In Sec.~\ref{frust_chain}, we present an alternative spin-interaction model for FeP and demonstrate its applicability by comparison to detailed neutron triple-axis spectroscopy (TAS) measurements. We discuss the obtained results and summarize the main findings of our study in Secs.~\ref{disc} and \ref{concl}, respectively.

\subsection{Magnetic structure of FeP}\label{magn_struc}

\begin{figure*}[t]
\includegraphics[width=0.99\linewidth]{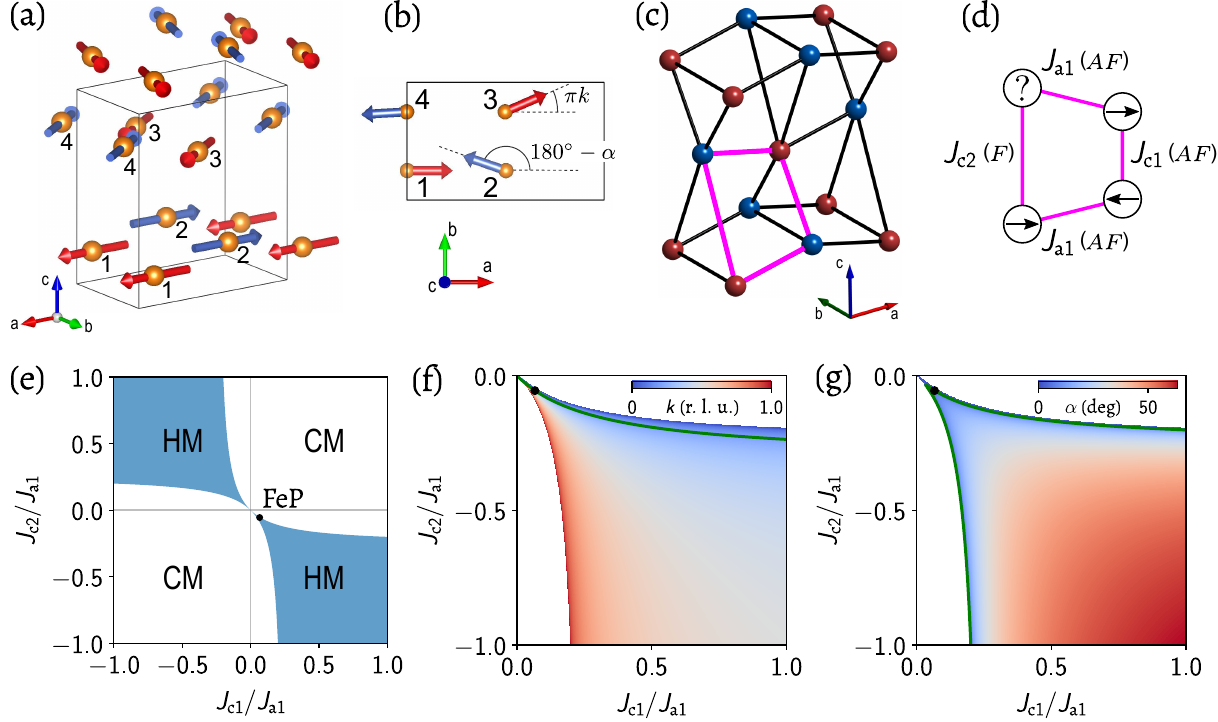}\vspace{3pt}
        \caption{(color online). (a) The double-helix magnetic structure of FeP reported in~\cite{Felcher71} (only Fe atoms are shown). The chosen enumeration of the Fe sites in the unit cell is shown beside the spins. (b) The same as in (a) but viewed from the $c$ axis. The spin angles are defined via the parameters $k$ and $\alpha$ (see text) as shown. (c) The Fe sublattice with the bonds between the nearest-neighbor sites shown as solid lines. A trapezoidal structural motif formed by four Fe sites in the unit cell is highlighted in magenta. (d) A schematic of the magnetic interactions induced within the trapezoid illustrating the magnetic frustration. (e) The classical phase diagram of the $J_{\text{a1}}$--$J_{\text{c1}}$--$J_{\text{c2}}$ model (the frustrated trapezoid, see text), HM and CM stand for the helimagnet and collinear magnet, respectively. FeP is placed on the phase diagram as suggested by~\cite{Kallel74}. (f) The absolute value of the lowest-energy propagation vector $k$ as a function of the exchange constants within the discussed model. (g) The lowest-energy canting angle $\alpha$ as a function of the exchange constants. The green solid lines show $k$ and $\alpha$ of FeP.}
        \label{ris:fig1}
\end{figure*}

The early neutron powder diffraction studies on FeP revealed that its magnetic structure below $T_{\text{N}} = 120$~K is not a simple spin spiral but a combination of two helices coupled nearly antiparallel to each other~\cite{Felcher71}. Such a type of magnetic structure was therefore named the double helix. The helices were found to propagate along the crystallographic $c$ axis (in the standard $Pnma$ setting), and the period of the spiral amounted to 29.2~\AA, which is close to the distance of 5 unit cells along $c$, yielding the propagation vector $\mathbf{k} = (0\,0\:0.2)$~ reciprocal-lattice units (r.l.u.), which are defined as $2\pi/a$, $2\pi/b$, and $2\pi/c$, where $a$, $b$, and $c$ are the real-space lattice parameters. The Fe ions were reported to host a weak magnetic moment of $\sim$0.4$\mu_{\text{B}}$~\cite{Felcher71}. A similar double-helical magnetic structure was found in the isostuctural compounds FeAs~\cite{Selte73,Rodriquez11,Frawley17}, MnP~\cite{Forsyth66,Moon82,Pan19}, and CrAs~\cite{Selte71,Matsuda18,Pan20}.

The unit cell of FeP has the lattice parameters $a = 5.197$, $b = 3.099$, and $c = 5.794$~\AA ~at room temperature, and contains four Fe ions at the Wyckoff site 4c. Their respective coordinates are $(x,1/4,z)$~(1), $(x+1/2,1/4,-z+1/2)$~(2), $(-x+1/2,3/4,z+1/2)$~(3), and $(-x,3/4,-z)$~(4), where $x = 0.002$ and $z = 0.200$. The P atoms also occupy the 4c position with the free parameters $x = 0.191$, $z = 0.569$~\cite{Felcher71}. If all the Fe atoms are projected onto the $c$ axis, their respective Fe-Fe distances along $c$ are 0.1$c$ for the pairs Fe(1)-Fe(2) and Fe(3)-Fe(4), and 0.4$c$ for the pair Fe(2)-Fe(3). Consequently, the distances Fe(1)-Fe(3) and Fe(2)-Fe(4) along the $c$ axis, which is the direction in which the spin spiral propagates, amount to 0.5$c$.

The chosen enumeration of the four Fe ions in the unit cell of FeP is illustrated in Fig.~\ref{ris:fig1}(a) along with its double helical magnetic structure. If the magnetic structure was a simple spin spiral, the propagation vector $\mathbf{k} = (0\,0\:0.2)$ would imply a spin rotation by 2$\pi k \times 0.5 = 36^{\circ}$ as one goes from Fe(1) to Fe(3), or from Fe(3) to Fe(1) in the next unit cell along the $c$ axis [colored as the red sublattice in Fig.~\ref{ris:fig1}(a)], as well as on the spins running along the sequence Fe(2)-Fe(4)-Fe(2) [colored as the blue sublattice in Fig.~\ref{ris:fig1}(a)]. The angle between the spins on the Fe(1) and Fe(2) sites would then be defined as 2$\pi k \times 0.1 = 7.2^{\circ}$ in a simple spin spiral. It was, however, found by Felcher \textit{et al.}~\cite{Felcher71} that the neutron diffraction data on FeP cannot be satisfactorily described without introducing a dephasing angle $\Delta \phi$ between the two sublattices. The nonzero dephasing angle means that in the rotating coordinate frame of the spin spiral, the spins on the two Fe sublattices are not parallel but disposed at an angle $\Delta \phi$ to each other. The neutron diffraction study of Ref.~\cite{Felcher71} provided $\Delta \phi = 168.8^{\circ}$. This corresponds to the relative angle of $176^{\circ}$ between the spins on the Fe(1)-Fe(2) bond in the basis of the unit cell. As we show in Fig.~\ref{ris:fig1}(b), it is convenient to introduce an angle $\alpha$ that measures the spin canting of the two sublattices away from the antiparallel orientation to describe the magnetic structure of FeP ($\alpha = 4^{\circ}$).

The details of the magnetic structure, including the value of the angle $\alpha$, were used in a number of studies to interpret the results of $^{57}$Fe M\"{o}ssbauer spectroscopy~\cite{Haggstrom82,Sobolev16}, $^{31}$P NMR spectroscopy~\cite{Sobolev16,Gippius20}, and the de Haas--van Alphen effect~\cite{Nozue01} measurements. Particularly, the authors of~\cite{Gippius20} showed that all the obtained $^{31}$P NMR spectra can be well reproduced theoretically when the parameter $\alpha = 4^{\circ}$ of the magnetic structure is taken into account, which further supports the neutron diffraction results~\cite{Felcher71}. The canting angle might play an important role in the unusual magnetotransport properties of FeP~\cite{Campbell21}, though the relation between the observed strongly anisotropic magnetoresistance and the distortions of the magnetic structure is rather complex and remains to be understood.

\subsection{Model of a frustrated trapezoid}\label{frust_trapez}

It is most important to understand how the double-helical magnetic structure of FeP is stabilized on the microscopic level. Figure~\ref{ris:fig1}(c) shows the crystal sublattice formed by the magnetic Fe ions (the P ions are omitted for clarity). The four Fe ions of the unit cell form a trapezoid that can be considered as a structural motif. The trapezoid has two equivalent nearest-neighbor bonds that are oriented in the $ac$ plane close to the $a$ axis and two bonds of slightly different length along the $c$ axis. We refer to the exchange interaction associated with the bond along $a$ as $J_{\text{a1}}$. Correspondingly, the two other exchanges are referred to as $J_{\text{c1}}$ and $J_{\text{c2}}$. As can be readily seen, if the two exchanges along $c$ are of the opposite sign, i.e. $J_{\text{c1}}J_{\text{c2}} < 0$, and in addition the spins are coupled along $a$ by a finite exchange (of any sign), then bond frustration is induced for spins residing on the vertices of the trapezoid. This relation is schematically shown in Fig.~\ref{ris:fig1}(d).

The model of the three nearest-neighbor Heisenberg exchange interactions in application to FeP and its related materials was first considered in the work of Kallel \textit{et al.}~\cite{Kallel74}. The results of their findings can be summarized as follows. The spin spiral with a propagation vector along $c$ minimizes the energy when the conditions $4J_{\text{c1}}J_{\text{c2}}/J_{\text{a1}}^2 + J_{\text{c1}}/J_{\text{a1}} + J_{\text{c2}}/J_{\text{a1}} < 0$ and $-4J_{\text{c1}}J_{\text{c2}}/J_{\text{a1}}^2 + J_{\text{c1}}/J_{\text{a1}} + J_{\text{c2}}/J_{\text{a1}} < 0$ are satisfied. The spiral pitch is then determined by the equation

\begin{equation}
\cos \pi k = -\frac{1}{4}\left( \frac{J_{\text{a1}}}{J_{\text{c1}}} + \frac{J_{\text{a1}}}{J_{\text{c2}}} \right).
\end{equation}
Otherwise, the collinear state is the ground state. The resulting phase diagram in the coordinates ($J_{\text{c1}}/J_{\text{a1}}$, $J_{\text{c2}}/J_{\text{a1}}$) is shown in Fig.~\ref{ris:fig1}(e). Each point on the phase diagram within the helical phase determines not only the magnitude of the propagation vector of the double spiral but also the canting angle $\alpha$ between the two spirals. It can be shown that the exchange parameters needed to minimize the spiral energy for the given values of $k$ and $\alpha$ can be expressed as~\cite{Kallel74}

\begin{equation}
\begin{split}
\frac{J_{\text{c1}}}{J_{\text{a1}}} = -\frac{1}{2} \frac{\sin\alpha}{\sin\left(\alpha - \pi k \right)}, \\
\frac{J_{\text{c2}}}{J_{\text{a1}}} = -\frac{1}{2} \frac{\sin\alpha}{\sin\left(\alpha + \pi k \right)}.
\end{split}
\label{eq:eq2}
\end{equation}
That means, the case of FeP ($k = 0.2$, $\alpha = 4^{\circ}$) in this exchange model corresponds to $J_{\text{c1}} = 0.066J_{\text{a1}}$ and $J_{\text{c2}} = -0.055J_{\text{a1}}$, as highlighted in Fig.~\ref{ris:fig1}(e).

It should be noted that there exists a set of points on the ($J_{\text{c1}}/J_{\text{a1}}$, $J_{\text{c2}}/J_{\text{a1}}$) plane that minimize the energy with respect to the same propagation vector (for any $\alpha$). This set of points forms a continuous curve on the phase diagram as demonstrated in Fig.~\ref{ris:fig1}(f). The same takes place for the energy minima for a fixed $\alpha$ [Fig.~\ref{ris:fig1}(g)], as there is a curve on the phase diagram corresponding to each stable $\alpha$. As can be seen, the curves intersect in only one point, which uniquely determines the ratio of the exchange parameters required for the double-spiral spin structure on the lattice built by the trapezoid motif.

In this paper, we closely examine the previously proposed model of a frustrated trapezoid and test its relevance to FeP. The model predicts that the exchange interactions in FeP should satisfy certain relations. This directly affects the magnon dispersions that can be probed by means of inelastic neutron scattering. A comparison between the observed and simulated spin-excitation spectra allows one to build an effective spin-interaction Hamiltonian capable of describing the material.

\section{Sample characterization and Experimental details}\label{exp_det}

\subsection{Sample description and characterization}

\begin{figure}[t]
        \begin{minipage}{0.99\linewidth}
        \center{\includegraphics[width=1\linewidth]{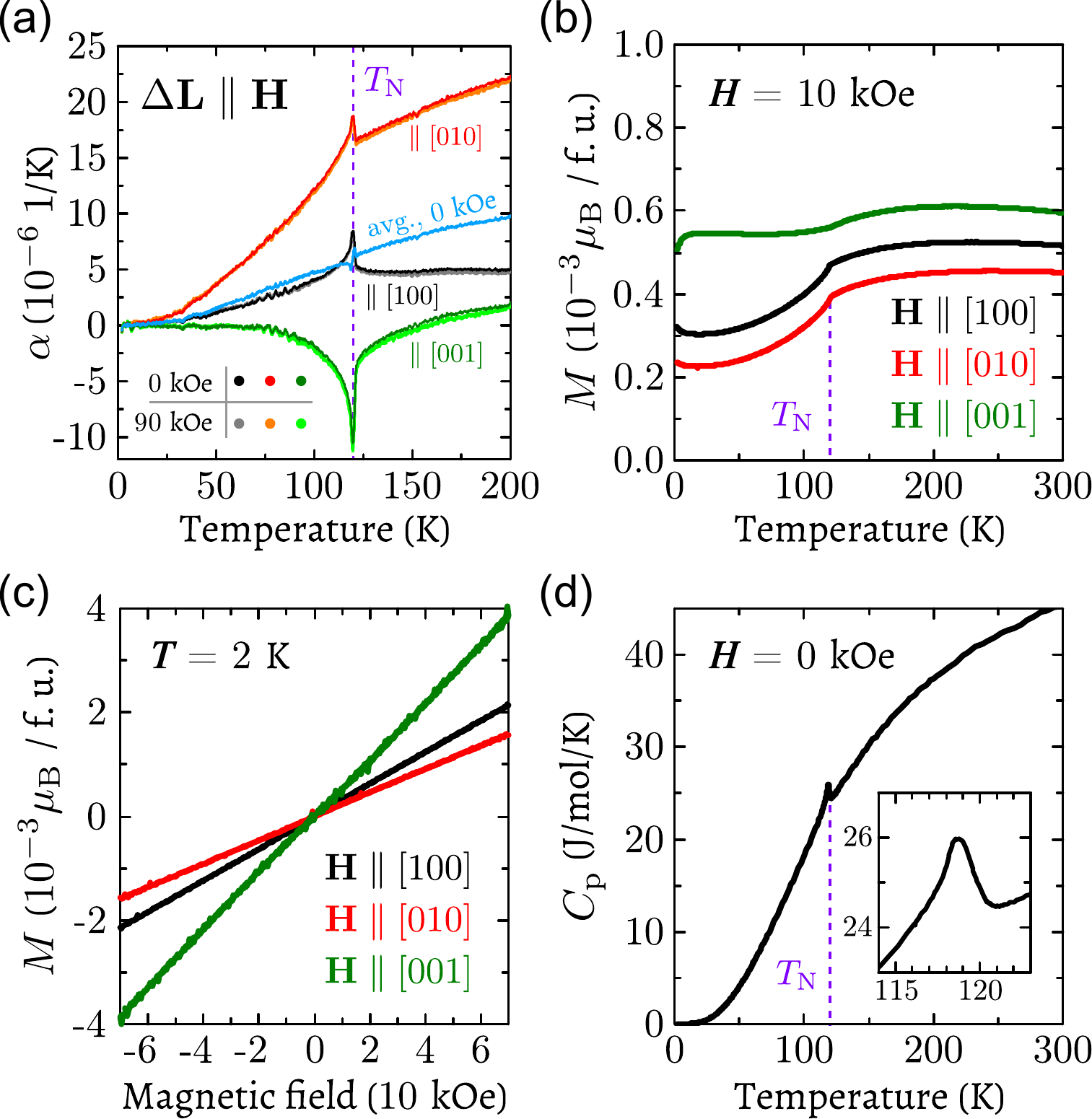}}
        \end{minipage}
        \caption{(color online). (a) Thermal expansion measurements of single-crystal FeP along the principal crystallographic directions in zero and applied magnetic fields. The blue line shows the average of three zero-field curves, which represents volume expansion. (b) Magnetization as a function of temperature for a magnetic field applied along the principal crystallographic directions. (c) Magnetization  curves at $T = 2$~K. (d) Specific heat measurements in zero field, with the transition region enlarged in the inset.}
        \label{ris:figS1}
\end{figure}

\begin{figure*}[t]
\includegraphics[width=0.99\linewidth]{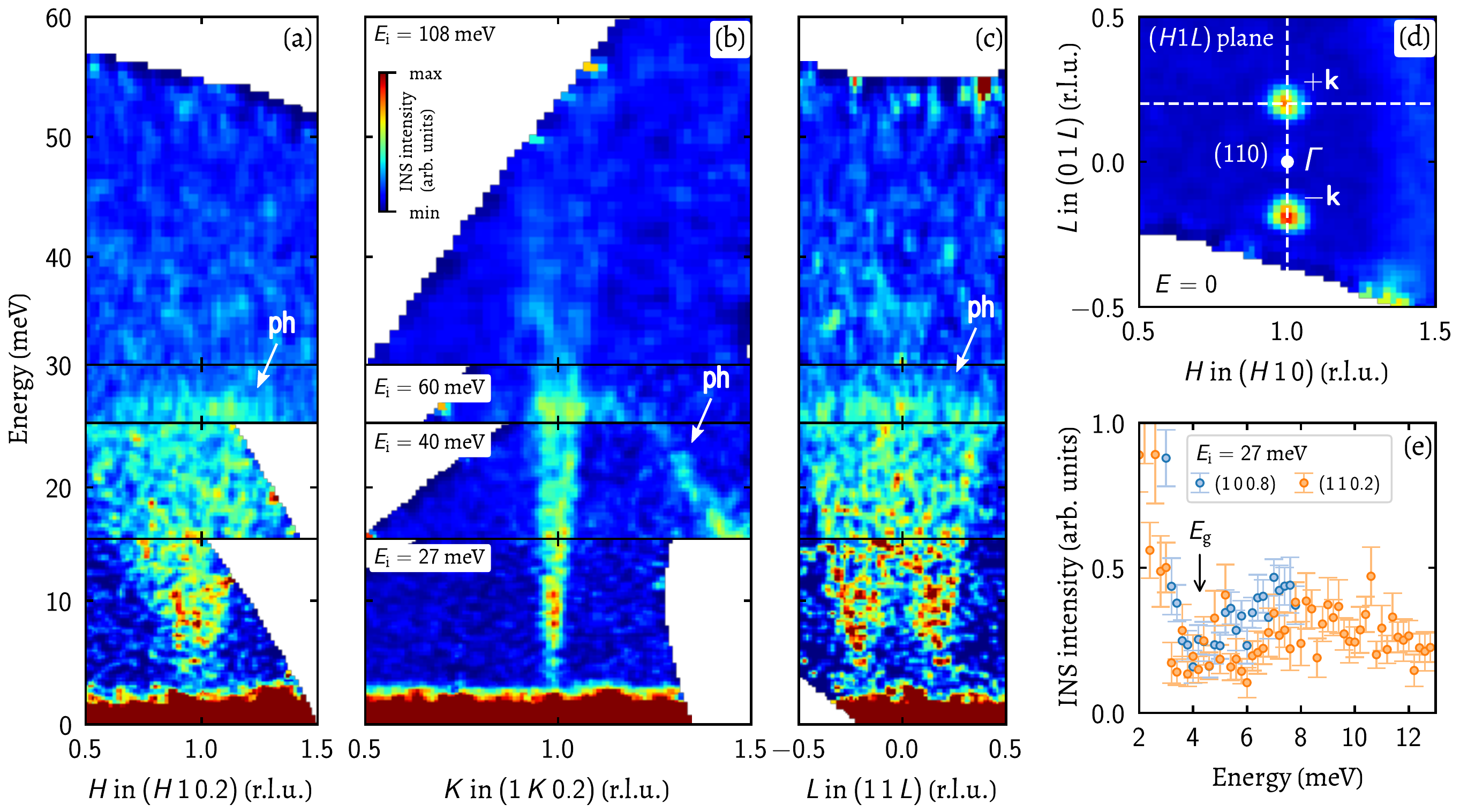}\vspace{3pt}
        \caption{(color online). Time-of-flight neutron spectroscopy data at $T = 7.5$~K. (a)--(c) The momentum-energy slice through the data for the $H$ (a), $K$ (b), and $L$ (c) reciprocal-space directions across the magnetic satellites of the (110) point. The data obtained with four different incident neutron energies $E_{\text{i}}$ were combined in a single plot as labelled. The data were integrated over a finite range in the perpendicular momenta (see Supplemental Materials~\cite{SM} for the integration range that corresponds to each slice). The white arrows show phonon modes. (d) The elastic scattering in the $(H1L)$ plane. The dotted lines show the direction of the momentum cuts for (a) and (c). (e) The INS intensity as a function of energy transfer for the momenta that correspond to the magnetic satellites of (101) and (110).}
        \label{ris:fig2}
\end{figure*}

Single crystals of FeP of high quality were grown using chemical vapor transport with an iodine transport agent \cite{Chernyavskii20}. The optimized temperature regime, reported in detail in Ref.~\cite{Chernyavskii20}, allowed us to grow large (up to 0.5\,g in mass) single crystals suitable for inelastic neutron scattering (INS) experiments. The grown crystals were extensively characterized to confirm their composition, crystal symmetry, unit cell parameters, and magnetic and transport properties in the earlier study~\cite{Chernyavskii20}. The elemental analysis and chemical characterization, done by means of energy-dispersive x-ray spectroscopy (EDX), revealed iron and phosphorus in 1:1 proportion corresponding to the FeP stoichiometric composition.

The single crystals used in the present study were also characterized by magnetization, dilatometry, and heat capacity measurements, as shown in Figs.~\ref{ris:figS1}(a)--\ref{ris:figS1}(d).

The magnetic transition in FeP can be characterized by measurements of the anisotropic thermal expansion coefficient $\alpha$ (see Supplemental Materials~\cite{SM} for the definition of $\alpha$ and details of the dilatometry measurements). Figure~\ref{ris:figS1}(a) shows $\alpha\left( T \right)$ from 2 to 200~K for all principal crystallographic directions. The magnetic transition is manifested by a sharp peak in $\alpha$ for the expansion along [010] and [100], and a dip for the expansion along [001]. Thus, the onset of the magnetic order leads to a pronounced anisotropic spontaneous magnetostriction. Below $T_{\text{N}}$, the unit cell of FeP is rapidly contracted in the $ab$ plane and expanded along the $c$ axis, such that the unit cell volume preserves a smooth change, as can be seen in the averaged thermal expansion curve plotted in Fig.~\ref{ris:figS1}(a). The examination of $\alpha(T)$ in a magnetic field of 9~T applied along $\Delta L/L$ did not reveal any noticeable induced magnetostriction, as also demonstrated in Fig.~\ref{ris:figS1}(a). The anisotropy of the spontaneous magnetostriction with respect to the crystallographic (001) plane matches the orientation of the spin spiral.

Figure~\ref{ris:figS1}(b) presents the magnetization in a field of 1~T as a function of the sample temperature measured on cooling from 300~K down to 2~K. The magnetization shows a very weak temperature dependence for the magnetic field applied along the three principle crystallographic directions until $\sim$120~K, where a clear downturn takes place, which indicates the onset of the antiferromagnetic ordering temperature $T_{\text{N}}$. The downturn in the $M(T)$ data is more apparent when the field is directed along either the [100] or [010] axis, whereas the field along [001] causes only a minor step in the magnetization at $T_{\text{N}}$. The following relation holds for the magnetization in the whole temperature range. The magnetization is higher for $H \parallel$~[001] and lower for $H \parallel$~[010]. It takes intermediate values when the field is applied along [100]. This is found to be in full agreement with previous reports~\cite{Chernyavskii20,Westerstrandh77}.

The $M(H)$ measurements are shown in Fig.~\ref{ris:figS1}(c). The magnetization at $T = 2$~K exhibits a linear behavior in applied magnetic fields up to at least 7~T, which is typical for an antiferromagnet. The slope of the $M(H)$ curve for $H \parallel$~[001] is nearly two times steeper than for $H \parallel$~[100] and $H \parallel$~[010]. As can be seen, the magnetic structure of FeP is mainly anisotropic with respect to the (001) plane, which is the spin-rotation plane of the helical structure. The magnetic measurements agree with the previous report~\cite{Westerstrandh77}.

Figure~\ref{ris:figS1}(d) shows the specific heat measurements in zero field. The N{\'e}el temperature is evidenced by a peak at $\sim$118~K, in agreement with the previous report~\cite{Nozue01}.

\subsection{Experimental configurations}

The time-of-flight (TOF) INS experiments were performed at the MERLIN spectrometer at ISIS, UK~\cite{Merlindoi}. A single crystal of FeP with a mass $\sim$0.5\,g was mounted in the top-loading closed-cycle cryostat with its $a$ axis vertical, providing the $(0\,K\,L)$ scattering plane. We collected the data at the base temperature of 7.5\,K using neutrons with incoming energies $E_\text{i} = $ 27, 40, 60, and 108\,meV. All the data processing was done using \textsc{Horace} software package \cite{Ewings16}.

Thermal-neutron TAS-measurements were performed at the IN8 (ILL, Grenoble)~\cite{IN8doi1,IN8doi2} and PUMA (MLZ, Munich)~\cite{PUMAdoi} spectrometers. All the TAS experiments were performed on a mosaic of FeP single crystals with a total mass $\sim$1\,g, coaligned with a backscattering x-ray Laue camera. During the IN8 measurements, the sample was mounted inside an ``orange''-type cryostat in either the $(H\,H\,L)$ or $(0\,K\,L)$ scattering plane. The experimental configuration with pyrolytic graphite (PG) filter and fixed final neutron wavenumber $k_\text{f} = 2.662$\,\AA$^{-1}$ was used to achieve a sufficient resolution for the measurements up to 35~meV energy transfer. To reach higher energies, we used $k_\text{f}$ set to 4.1\,\AA$^{-\!1}$. Measurements were done at the base temperature of 2\,K.
 
In the PUMA experiment, the sample was mounted in the standard closed-cycle cryostat with its $b$ axis vertical, giving access to the $(H\,0\,L)$ scattering plane. The measurements were performed at the temperature of 3.5\,K. To reach a compromise between intensity and resolution, the instrument was operated with the PG(002) monochromator and analyzer in the double-focusing mode. Measurements were performed with fixed $k_\text{f} = 2.662$\,\AA$^{-\,1}$ up to 15 and 12\,meV for $L$ and $H$ scans, respectively.
To obtain sufficient momentum coverage for higher energies, we changed $k_\text{f}$ to $4.1$\,\AA$^{-\!1}$, which naturally entails lower resolution.  

\section{TOF spectroscopy}\label{TOF}

\subsection{Observed spectra}\label{tof.observed}

In order to discuss the general characteristics of the magnetic excitations in FeP, we begin with a presentation of the low-temperature TOF data that covers a large part of the 4D momentum-energy space. Such an approach allows one to identify the reciprocal-lattice planes at which the significant magnon spectral weight can be observed by thorough analysis of the individual slices through the collected dataset. Moreover, the momentum-energy $(Q$-$E)$ slices for different crystallographic directions reveal the magnon-disperion bandwidth along all the high-symmetry paths in the first Brillouin zone (BZ). Note that the magnetic sublattice of iron atoms in FeP has a higher symmetry than the lattice itself, therefore spin-wave dispersions can be described in either the folded or unfolded BZ notation, as explained in the Supplemental Materials~\cite{SM}.

Figures~\ref{ris:fig2}(a)--\ref{ris:fig2}(c) summarize the main features of the magnetic spectra of FeP at a temperature of 7.5~K. The previous neutron diffraction measurements on single crystals~\cite{Felcher71,Chernyavskii20}, as well as our powder neutron diffraction data shown in the Supplemental Materials~\cite{SM} showed that the helical magnetic structure yields pairs of strong magnetic satellites at the momenta $(1\,1\,0) \pm\!\mathbf{k}$ and $(1\,0\,1) \pm\!\mathbf{k}$, where $\mathbf{k} = (0\,0\,\:0.2)$~r.l.u. The elastic ($E = 0$) slice through the dataset in the vicinity of the momentum transfer $\mathbf{Q} = (1\,1\,0)$, which is a forbidden Bragg reflection for the nuclear structure, is shown in Fig.~\ref{ris:fig2}(d). The magnetic Bragg peaks are observed at $(1\,1\,\pm 0.2)$, which yields $\mathbf{k} = (0\,0\,\:0.2)$ in agreement with the previous reports~\cite{Felcher71,Chernyavskii20}. Figures~\ref{ris:fig2}(a)--\ref{ris:fig2}(c) present the $(Q$-$E)$ slices in a wide range of energy transfer up to 60~meV~for the reciprocal-space directions $H$, $K$, and $L$, respectively, all intersecting the magnetic Bragg peak [as illustrated in Fig.~\ref{ris:fig2}(d)]. The covered energy range ($\approx$~700~K~$\approx$~6$T_{\text{N}}$) significantly exceeds the energy scale set by the magnetic ordering temperature. The data were collected at different energies of the incident neutrons $E_{\text{i}}$ to combine a large accessible energy-transfer range at higher $E_{\text{i}}$ with an improved resolution at lower incident energies. The resulting datasets were composed in Figs.~\ref{ris:fig2}(a)--\ref{ris:fig2}(c) for completeness.

As can be seen, the magnon dispersion along $L$ [Fig.~\ref{ris:fig2}(c)] consists of two V-shaped branches stemming from the momentum that corresponds to the two magnetic satellites. The dispersions intersect at $\mathbf{Q} = (1\,1\,0)$, which is the center of the BZ (the $\Gamma$ point), at $\sim$15~meV forming a characteristic W-shaped spectrum.

The magnon dispersion reaches an energy of $\sim$25~meV at the BZ boundary at $\mathbf{Q} = (1\,1\,\pm\!0.5)$, where a weakly dispersing intense optic-phonon mode is observed above the magnon band. The phonon mode disperses weakly also for the $H$ direction, but has a pronounced dispersion down to $\sim$15~meV along $K$. The assessment of the phonon mode as an optic phonon is based on the previous detailed study of the lattice vibrations in FeP~\cite{Sukhanov20}. The fact that the magnon bandwidth along the $L$ direction in momentum space is bounded by $E = 25$~meV is confirmed by the absence of any INS intensity at higher energies at least up to $\sim$55~meV covered in the collected dataset. 

The magnetic dispersion along the orthogonal $H$ direction is shown in Fig.~\ref{ris:fig2}(a), where the INS intensity is plotted for the momenta $(H\,1\,0.2)$ in a wide energy range. The magnon dispersion, approximately linear in the vicinity of the magnetic Bragg peak, also reaches up to $\sim$25~meV at the BZ boundary. Thus, the magnon bandwidths for the dispersions along $L$ and $H$ are the same within our experimental resolution. Similarly to the ($Q$-$E$) slice along $L$, the data above $E = 25$~meV along $H$ did not reveal any additional intensity.

Remarkably, the spin-wave dispersion of FeP along the reciprocal-space direction $K$ exhibits a striking difference to that of the $H$ and $L$ directions. As can be seen in Fig.~\ref{ris:fig2}(b), the linear dispersion already acquires an energy of 50~meV at nearly 0.075 r.l.u. from (1\,1\,0.2) and far away from the BZ boundary [$(1\:1\!\pm\!0.5\;0.2)$]. This yields a spin-wave stiffness nearly ten times higher along $K$ than in the $H$ and $L$ directions. Such a large anisotropy in the magnetic exchange energies along different crystallographic directions is unusual for a structurally 3D material, as it is more common for compounds with isolated chains of magnetic ions (typically with the atomic distance within the chains being much smaller than the distance between the adjacent chains)~\cite{Coldea96,Songvilay21,Mourigal13,Enderle05}. A crude extrapolation of the magnon dispersion along $K$ above the measured energy range suggests that the energy scale of the exchange interactions along the $b$ axis in FeP exceeds $T_{\text{N}}$ by a factor of $\sim$30, which explains why magnetic correlations are seen in the magnetic susceptibility measurements up to 500~K~\cite{Westerstrandh77} [also supported by the broad maximum in our data up to 300~K in Fig.~\ref{ris:figS1}(b)].

The TOF data shown in Figs.~\ref{ris:fig2}(a)--\ref{ris:fig2}(c) reveal also that the magnetic excitations in FeP are gapped. This means that there exists a non-negligible single-ion magnetic anisotropy, which should be taken into account when constructing a relevant magnetic Hamiltonian capable of describing the spin dynamics of FeP. To further illustrate this, we plotted the INS intensity as a function of energy at the magnetic Bragg peak positions, $(1\:1\:0\pm\!0.2)$, and at $(1\:0\:1\!\pm\!0.2)$ [Fig.~\ref{ris:fig2}(e)]. For this, the data for the $\pm \mathbf{k}$ cuts were averaged to increase the statistics. The intensity profiles in Fig.~\ref{ris:fig2}(e) show a clear minimum at $\sim$4.5~meV, labelled as $E_{\text{g}}$, at both equivalent momenta. The value of $E_{\text{g}}$, however, is not directly equal to the magnon gap, as the position of the minimum in the intensity profile is affected by finite resolution effects for the out-of-plane momenta.

It should be noted that the spectra in Figs.~\ref{ris:fig2}(a) and \ref{ris:fig2}(c) are not seen as sharp modes but rather appear as a continuous intensity distribution that fills energies above the magnon branches. This can be explained by the finite momentum resolution in the $K$ direction. Because the dispersion along $K$ is very steep, it visibly smears the spectra for $H$ and $L$. This should be taken into account when the experimental spectra are compared to the simulated ones.

\subsection{Comparison to the $J_{\text{a1}}$--$J_{\text{c1}}$--$J_{\text{c2}}$ model}

\begin{figure*}[t]
\includegraphics[width=0.9\linewidth]{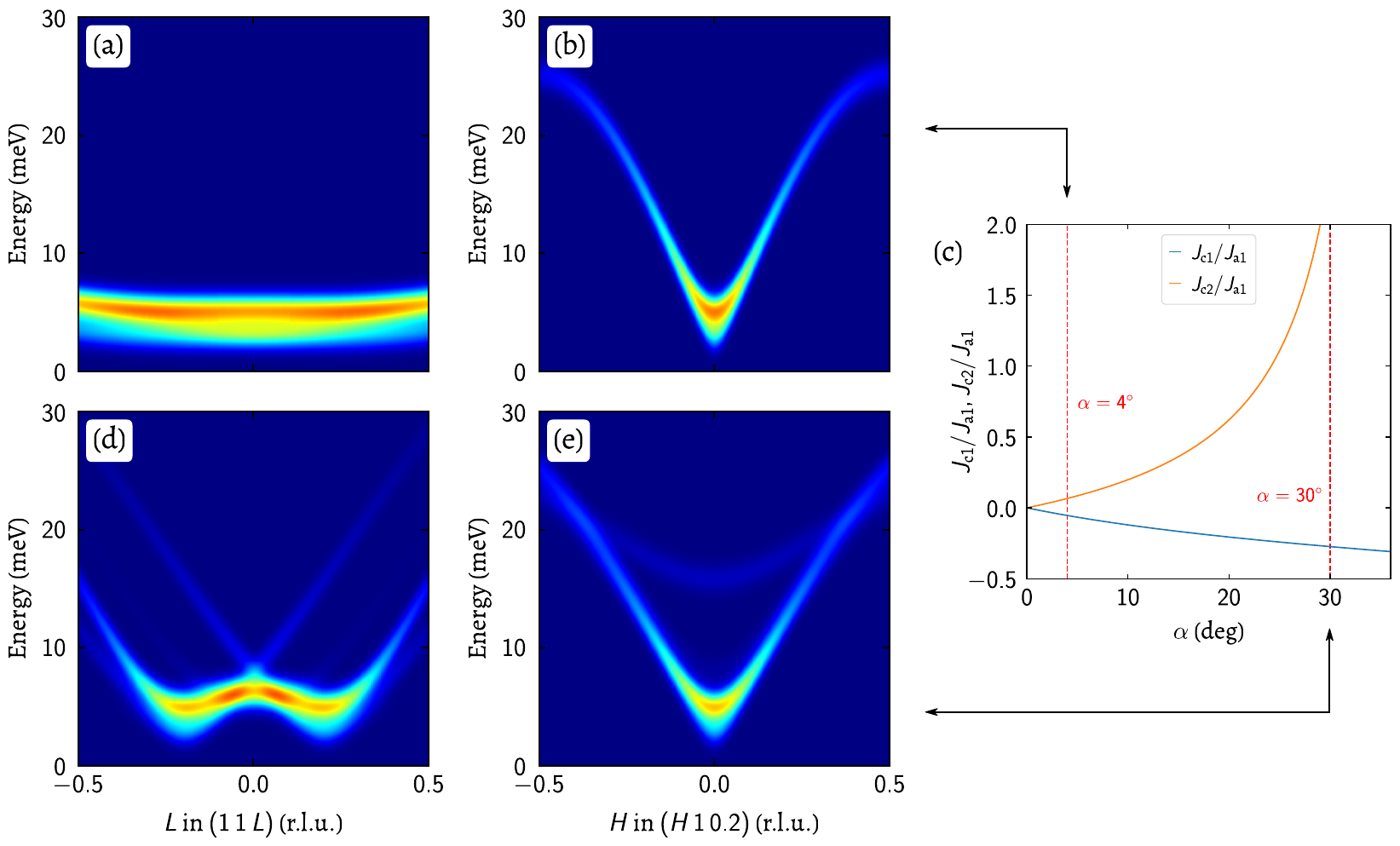}\vspace{3pt}
        \caption{(color online). The simulated magnon spectra for FeP within the model of frustrated trapezoid along the $L$~(a,d) and $H$~(b,e) reciprocal-space directions for the parameter $\alpha = 4^{\circ}$ (a,b) and $\alpha = 30^{\circ}$ (d,e). (c) The ratios of the exchange interactions $J_{\text{c1}}/J_{\text{a1}}$ and $J_{\text{c2}}/J_{\text{a1}}$ as a function of the angle $\alpha$ highlighting the case of $\alpha = 4^{\circ}$ and $\alpha = 30^{\circ}$.}
        \label{ris:fig3}
\end{figure*}

Having observed the magnon dispersions along all the high-symmetry directions in reciprocal space, we can now compare them to the predicted excitation spectra of the frustrated trapezoid model~\cite{Kallel74}. To be more specific, only spectra for the $H$ and $L$ directions become important for this comparison, whereas the dispersion along $K$ does not depend on the parameters of the discussed model and can be omitted in the analysis. Indeed, the frustrated exchange interactions $J_{\text{a1}}$, $J_{\text{c1}}$, and $J_{\text{c2}}$ link the neighboring spins within the $ac$ plane of the crystal structure, which corresponds to the $(H0L)$ plane in reciprocal space. The magnon dispersion along $K$, in turn, is solely driven by the exchange interaction that couples the spins exactly along the $b$ axis, which can be labelled as $J_{\text{b}}$.

The ground state of FeP clearly suggests that $J_{\text{b}}$ is ferromagnetic (FM). Furthermore, as one can conclude from the very steep magnon dispersion along $K$, $J_{\text{b}}$ must be much larger in absolute value than the AFM or FM exchanges along $a$ and $c$, $|J_{\text{b}}| \gg J_{\text{a,c}}$. In other words, the magnetic structure of FeP can be viewed as strongly-coupled FM chains running along the crystallographic $b$ axis. The chains are coupled in the perpendicular $ac$ plane with weaker AFM or FM interactions. The double-helical spin structure of FeP is thus stabilized on a relatively low energy scale.

Figures~\ref{ris:fig3}(a) and \ref{ris:fig3}(b) show the simulated dispersion along $L$ and $H$, respectively. The simulations were performed within the linear spin-wave theory as implemented in \textsc{SpinW}~\cite{Toth2015}. The exchange parameters used for these simulations corresponds to the ratio of $J_{\text{c1}}/J_{\text{a1}}$ and $J_{\text{c2}}/J_{\text{a1}}$ that stabilize the magnitude of the propagation vector $k = 0.2$ and the canting angle $\alpha = 4^{\circ}$ as previously reported~\cite{Felcher71}. Figure~\ref{ris:fig3}(c) demonstrates the calculated exchanges as a function of $\alpha$, as defined by Eq.~(\ref{eq:eq2}), for the fixed value of $k = 0.2$. It is easy to see that the simulated spectra along $L$ [Fig.~\ref{ris:fig3}(a)] drastically disagree with the experimental observations [Fig.~\ref{ris:fig2}(c)]. Whilst the simulated dispersion along $H$ can be brought into agreement with the experimental data by tuning $J_{\text{a1}}$ to the observed bandwidth of $\sim$25~meV, the dispersion along $L$ appears to be completely flat. This result is not surprising, as $J_{\text{c1}}/J_{\text{a1}}$ and $J_{\text{c2}}/J_{\text{a1}}$ for $\alpha = 4^{\circ}$ are fixed to very small values of 0.066 and $-0.055$, respectively. To simulate the spectra in Figs.~\ref{ris:fig3}(a) and \ref{ris:fig3}(b), we also added a finite single-ion anisotropy in the $ab$ plane to reproduce the observed spin gap for a better comparison.

To further examine the previously proposed model of the frustrated trapezoid, one can investigate how the simulated spectra depend on the canting angle $\alpha$ for the same propagation vector. It should be noted that, in principle, $\alpha$ only affects the relative intensities of the magnetic Bragg peaks in a neutron diffraction pattern. Thus it cannot always be experimentally extracted with a high confidence. This is in contrast to the propagation vector, which affects the position of the peaks and can therefore be well resolved.

As can be seen from Eq.~(\ref{eq:eq2}), the model predicts the limiting value of $\alpha$ at which the helical solution no longer exists, $\alpha < \pi k$, which becomes $\alpha < 36^{\circ}$ in the case of FeP. According to Eq.~(\ref{eq:eq2}), the ratio of $J_{\text{c2}}/J_{\text{a1}}$ changes rapidly as $\alpha$ increases [Fig.~\ref{ris:fig3}(c)]. The exchange $J_{\text{c2}}$ becomes equal to $J_{\text{a1}}$ at $\alpha \approx 25^{\circ}$. At $\alpha = 30^{\circ}$, $J_{\text{c2}}$ has already become twice larger than the exchange along $a$, and it then diverges as one approaches the maximal value of $\alpha$.

In order to test if the $J_{\text{a1}}$--$J_{\text{c1}}$--$J_{\text{c2}}$ model can describe the experimental data for a different ratio between the exchange interactions along $a$ and $c$, we simulated the spectra for a significantly larger $\alpha$ and plotted them in Figs.~\ref{ris:fig3}(d) and \ref{ris:fig3}(e) for the reciprocal $L$ and $H$ directions, respectively. Because at $\alpha = 30^{\circ}$ the exchange interactions along $a$ and $c$ are of the same order, the magnon dispersion along $L$ acquires a bandwidth that is comparable to the bandwidth along $H$. While in this aspect the simulated spectra become reminiscent to the observed ones, the spectrum in Fig.~\ref{ris:fig3}(d) is still much different from the data in Fig.~\ref{ris:fig2}(c) in other aspects. The simulated spectrum also consists of two V-shaped branches separated by $\pm k$, but they cross at a much lower energy of only $\sim$7~meV. The simulated spectrum also predicts the lower intense mode to reach an energy of $\sim$15~meV at the BZ boundary, whereas the experimental data suggests that the lower mode along $L$ disperses up to 25~meV. A closer examination of the model shows that the magnon bandwidth along $L$ and $H$ for the most intense lower band cannot be equalized for any $\alpha$ up to the maximal value of $36^{\circ}$. Therefore, one can conclude that the model proposed in~\cite{Kallel74} is not applicable for FeP.

\section{Model of frustrated chains}\label{frust_chain}

\subsection{Exchange scheme}

\begin{figure}[t]
        \begin{minipage}{0.99\linewidth}
        \center{\includegraphics[width=1\linewidth]{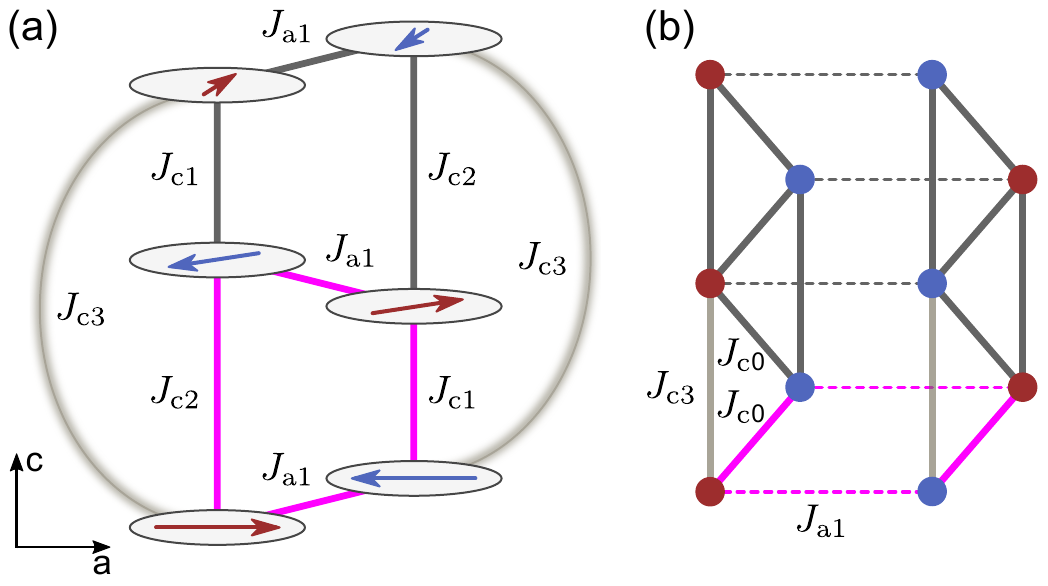}}
        \end{minipage}
        \caption{(color online). (a) The exchange interaction scheme of the model of frustrated AFM chains. The crystal-structure sublattice of Fe atoms is shown in the $ac$ plane. The magenta trapezoid highlights the structural motif formed by the four Fe atoms in the unit cell of FeP. (b) The schematics of the frustrated zigzag spin chains, which represent the same model as in (a) when the difference between the $J_{\text{c1}}$ and $J_{\text{c2}}$ bonds is neglected.}
        \label{ris:fig4}
\end{figure}

Since the model of a frustrated trapezoid~\cite{Kallel74} turns out to contradict the experimental data, we should consider an alternative mechanism capable of stabilizing the spin spiral in FeP. One such mechanism is the frustration between the nearest-neighbor and the next nearest-neighbor spins along the propagation direction of the spiral. Indeed, this type of frustration was discussed as the origin of the helical magnetic structure in a number of materials, for example in Fe$_3$Ga$_4$~\cite{Afshar21}, MnAu$_2$~\cite{Udvardi06,Glasbrenner14}, and YMn$_6$Sn$_6$~\cite{Ghimire20}, to name a few.

Figure~\ref{ris:fig4}(a) displays the exchange interaction scheme that we refer to as the model of frustrated chains, where the frustrating interactions are along the $c$ axis. We note that the term ``chains'' in this context is not to be confused with the rigid FM spin chains along the $b$ axis, which are confined by the dominating exchange as was discussed in Sec.~\ref{tof.observed}. Because this extremely strong exchange along $b$ effectively turns every FM chain into one large classical magnetic moment, the magnetic model becomes effectively 2D, confined to the $ac$ plane. We then use the term chains in a broad sense to refer to the spins connected by the bonds running along $c$ [in the sequences 1-4-1 and 2-3-2 according to Fig.~\ref{ris:fig1}(a)].

The frustration in this case can be explained as follows. Let the spins along $c$ axis couple by the same AFM exchange interaction, $J_{\text{c1}} = J_{\text{c2}} = J_{\text{c0}} > 0$. Without any further interaction, the ground state is an AFM order of the alternating up-down-up spins along $c$. If another AFM interaction, $J_{\text{c3}} > 0$, is introduced between the spins that are second neighbors along $c$, then an AFM collinear order twists into an AFM spiral. The AFM spirals on the adjacent spin chains running along the $c$ axis can be further coupled in an AFM fashion along $a$ by $J_{\text{a1}}$ and ferromagnetically along $b$ by $J_{\text{b}}$. The main difference of this model, as compared to the model of~\cite{Kallel74}, is that $J_{\text{a1}}$ no longer influences the spiral propagation vector, thus can be chosen independently from $J_{\text{c1}}$ and $J_{\text{c2}}$. It can be easily shown, that the spiral propagation vector is given by the simple relation:

\begin{equation}
\cos(\pi k) = -\frac{J_{\text{c0}}}{4J_{\text{c3}}}.
\label{eq:eq3}
\end{equation}
It is obvious that the exchange interaction along $b$ remains decoupled from the other exchange interactions in this model as it was in the model of the frustrated trapezoid. The apparent drawback of the model of frustrated chains (also referred to as the $J_{\text{c0}}$--$J_{\text{c3}}$ model) is that the canting angle between the adjacent spin spirals $\alpha$ always remains zero and cannot be driven away from that value. This, however, can be considered as a minor approximation, as the reported $\alpha = 4^{\circ}$~\cite{Felcher71} is very small.

The model in which the spins are frustrated by the competition of the nearest- and the next-nearest-neighbor AFM exchange interactions is also known as the zigzag spin chain model, which was widely studied in the quantum limit~\cite{White96}. Figure~\ref{ris:fig4}(b) illustrates the analogy between the zigzag chains defined by the $J_{\text{c0}}$ and $J_{\text{c3}}$ interaction and the original structural motif of the Fe sublattice in FeP. We note that we consider the model only in the classical limit.

\subsection{TAS measurements}

\begin{figure}
        \begin{minipage}{0.95\linewidth}
        \center{\includegraphics[width=1\linewidth]{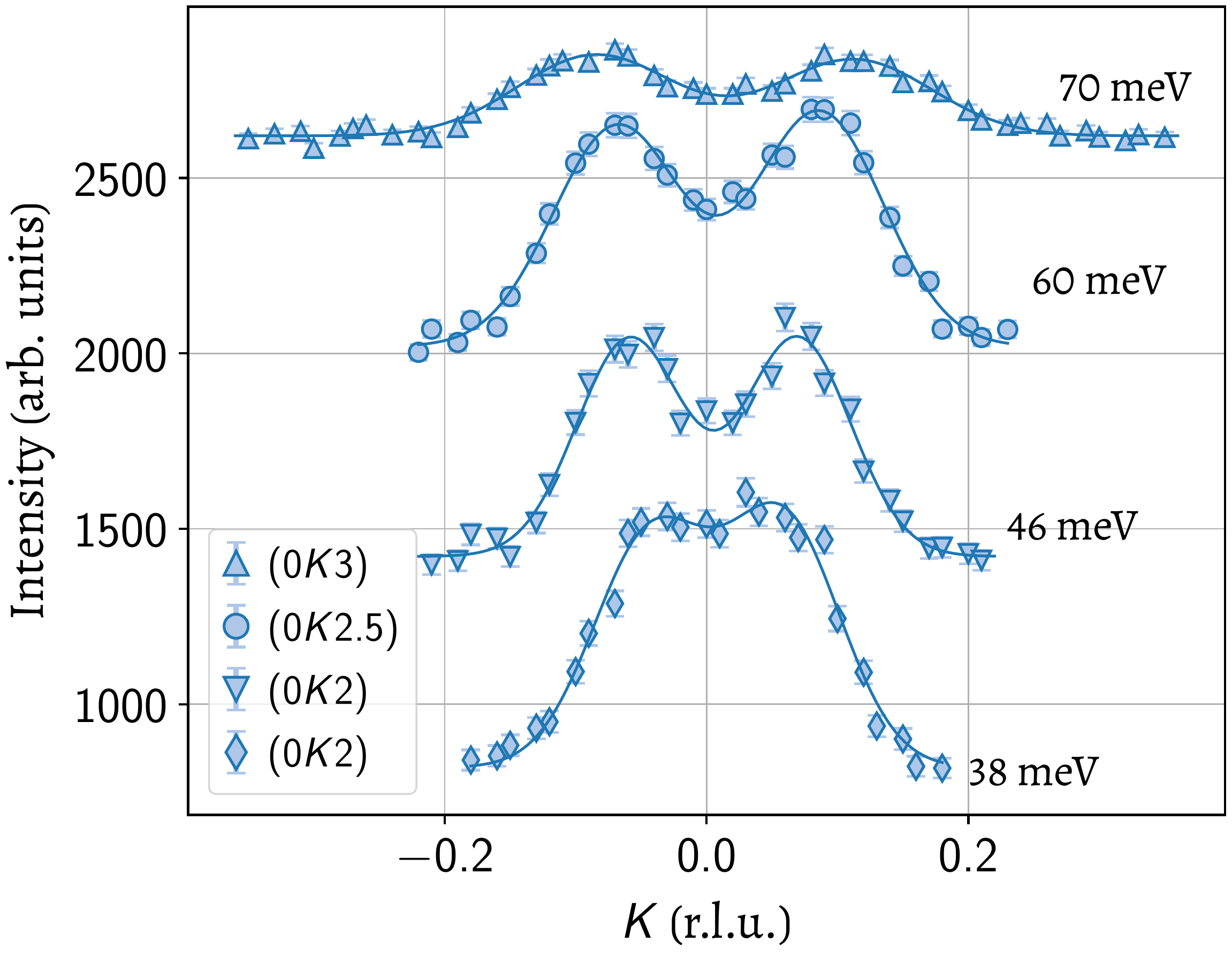}}
        \end{minipage}
        \caption{(color online). TAS constant-energy data (symbols) for the measurements along the $K$ direction in the $(0KL)$ plane shown for different energies and different values of $L$. The solid lines are fits by a sum of two Gaussian functions. The data were offset for clarity.}
        \label{ris:figA2}
\end{figure}

The TOF data presented above enabled us to understand the main aspects of the spin excitation spectra of FeP. For a thorough comparison with the relevant exchange interaction model, however, fine details of the spectra for the reciprocal $H$ and $L$ directions are essential. This was achieved by performing triple-axis spectroscopy measurements. We mapped out the magnon dispersions by performing constant-energy scans for the selected directions in momentum space, namely along $L$ at the (110), (101), and (002) zones and along $H$ at the (101) zone. The measurements in the (110) zone thus reproduce the results obtained with TOF but with significantly better statistics, yet with a similar energy resolution. As a result of these measurements, subtle details of the excitation spectra were revealed.

Before further discussing the magnon dispersions along $H$ and $L$, we verify that the magnon dispersion is very steep along $K$ beyond the energy transfer range covered in our TOF data (Sec.~\ref{TOF}). Performing TAS measurements in the $(0\,K\,L)$ scattering plane along $K$ at $(00L)$ with $L=2, 2.5, 3$ with tight collimation to improve momentum resolution, we could resolve pairs of peaks along the $K$ direction up to an energy of 70 meV, as shown in Fig.~\ref{ris:figA2}. At the maximal energy transfer of 70~meV, the peaks are centered at approximately $\pm$0.1 r.l.u., which confirms that the magnon bandwidth along the $b$ direction must be one order of magnitude larger than along $a$ and $c$.

Figures~\ref{ris:fig5}(a) and ~\ref{ris:fig5}(b) show the measurements along $L$ performed in the vicinity of the equivalent reciprocal-space points  (110) and (101). The observed spectra look very similar in terms of the resolved magnon dispersions, while the spectral weight inherits the intensities of the underlying magnetic Bragg peaks that are equal for $(110)\pm\mathbf{k}$ by symmetry but differ for $(101)\pm\mathbf{k}$ due to the structure factor. It should be noted that the weakly-dispersing mode at $E \approx 27$~meV in the ($11L$) data corresponds to an optic phonon, which was also identified in the TOF data discussed above.

Figure~\ref{ris:fig5}(c) demonstrates the measurements done with longitudinal scans  along $L$ in the (002) zone covered up to a lower energy cut-off of $E = 15$~meV. It is nevertheless very useful to compare it to the data at the (110) and (101) zones. Again, here the structure factor leads to different intensities of the spin-wave branches emanating from the nonequivalent magnetic Bragg peaks. The $(002)$ data differ from those taken in the other two zones in one important aspect. As one can see from the transverse or mixed scans in Figs.~\ref{ris:fig5}(a) and \ref{ris:fig5}(b), the INS intensity at the $\Gamma$ point appears at $\sim$10~meV, which is much lower than the energy at which the two $V$-shaped modes apparently cross ($\sim$18~meV). This indicates that there is a low-energy mode that connects the two minima at $\pm \mathbf{k}$ via a $\cap$-shaped dispersion. The absence of INS intensity up to $\sim$15~meV in Fig.~\ref{ris:fig5}(c) suggest that this mode has zero or nearly vanishing spectral weight at the (002) zone. This, in turn, implies that the $\cap$-shaped branch corresponds to spin fluctuations polarized along the $c$ axis, as these would vanish along $(00L)$ due to the neutron polarization factor.

The TAS data for the $H$ direction are shown in Fig.~\ref{ris:fig5}(d). They clearly reveal the dispersion of the lowest magnon branch by a relatively sharp onset of the INS intensity. Above this boundary line, the intensity looks continuous due to a finite momentum resolution in the $K$ direction orthogonal to the $(H0L)$ scattering plane, in which the magnon dispersion along $K$ is very steep [see Fig.~\ref{ris:fig2}(b) and \ref{ris:figA2}]. The maximal INS intensity at the zone boundary along $H$ is found at $\sim$25~meV, in agreement with the TOF measurements [Fig.~\ref{ris:fig2}(a)].

\subsection{Comparison to the $J_{\text{c0}}$--$J_{\text{c3}}$ model}

\begin{figure*}[t]
\includegraphics[width=0.99\linewidth]{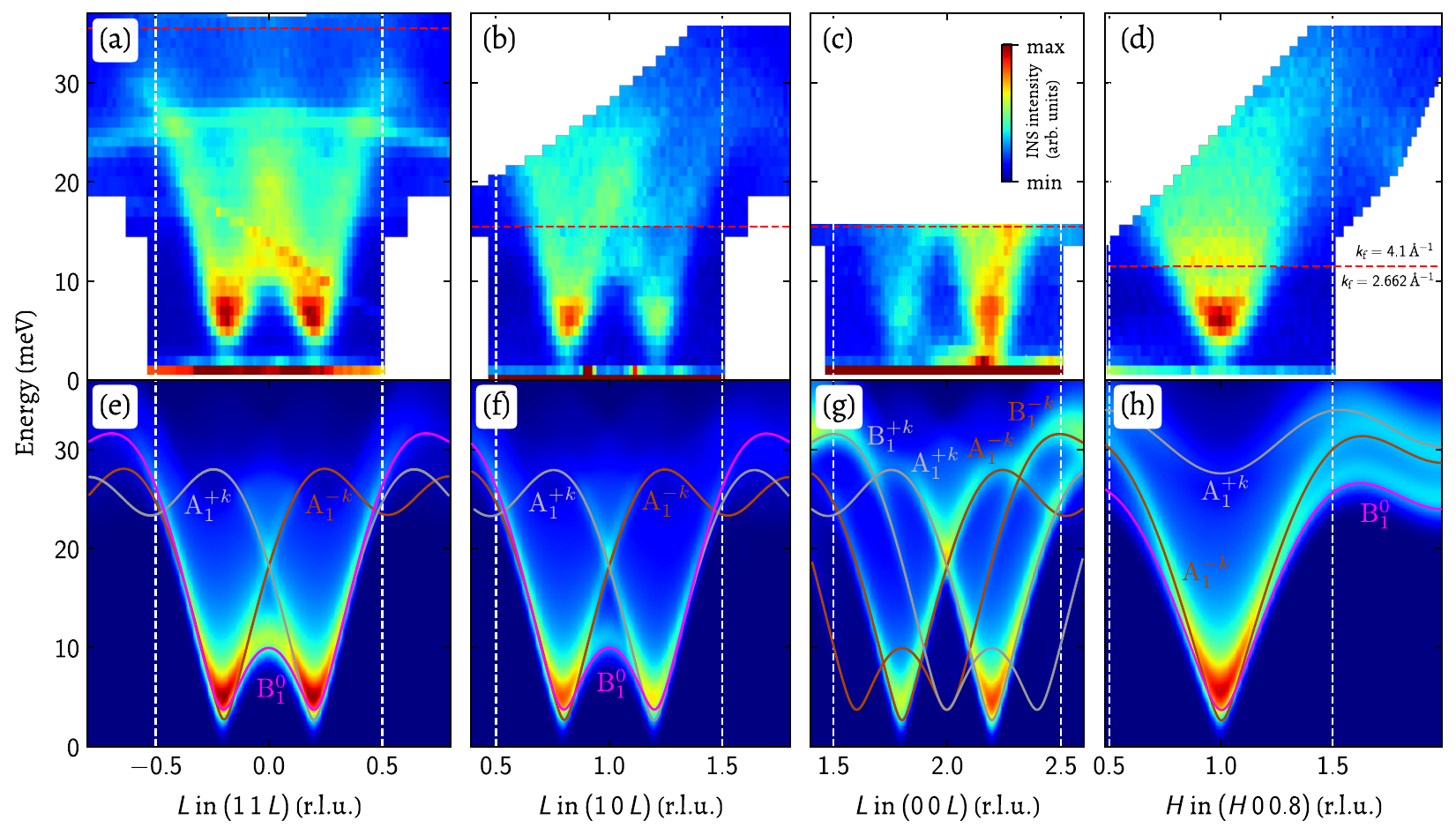}\vspace{3pt}
        \caption{(color online). (a)--(d) The TAS measurements of FeP for the momenta along the reciprocal directions $(11L)$ at $T = 2$~K (a), $(10L)$ at $T = 3.5$~K (b), $(00L)$ at $T = 2$~K (c), and $(H\,0\,0.8)$ at $T = 3.5$~K (d). The horizontal red dotted lines separate the data collected with different $k_f$. The sharp streak of intensity in the $\sim$5--20~meV range in (a) is a measurement artefact (due to neutrons that scattered at the (111) Bragg peak of the sample and further scattered incoherently off the analyser). The spectrally-sharp weakly dispersing mode at $\sim$25~meV is an optic phonon~\cite{Sukhanov20}. (e)--(h) The linear spin-wave theory simulations of the magnon spectra (colormap) in FeP within the model of Eq.~(\ref{eq4}). To account for the energy-momentum resolution in the TAS measurements, all the simulated spectra were broadened by 3~meV in energy and integrated in the perpendicular momenta over $\pm 0.06$~r.l.u. in $H$ and $\pm 0.04$~r.l.u. in $K$ in (e)--(g), and $\pm 0.07$~r.l.u. in $L$ and $\pm 0.04$~r.l.u. in $K$ in (h). The solid lines show the simulated dispersion relations (only the modes that show a significant INS intensity are drawn, for the full set of dispersions see Supplemental Materials~\cite{SM}). Vertical dashed white lines mark the orthorhombic BZ boundary.}
        \label{ris:fig5}
\end{figure*}

We now can compare the observed TAS spectra with the magnon spectra simulated within the $J_{\text{c0}}$--$J_{\text{c3}}$ model (the model of frustrated chains or zigzag chains). The full Hamiltonian used for the simulations can be written as
\begin{equation}\label{eq4}
\begin{split}
\mathcal{H} = \sum_{ijk} S_{ijk} \left(  S_{\{i+1\}jk} J_{\text{a}1} + S_{\{i+2\}jk} J_{\text{a}2} + S_{i\{j+1\}k} J_{\text{b}} \right. + \\ + \left. S_{ij\{k+1\}} J_{\text{c}0} + S_{ij\{k+2\}} J_{\text{c}3}  + A S_{ijk} \right),
\end{split}
\end{equation}
where the indexes $i$, $j$, and $k$ enumerate the spins along the crystal-lattice $a$, $b$, and $c$ axes, respectively. The single-ion anisotropy matrix $A$ describes the orientation of the easy plane and is essential for the model to reproduce the observed spin-wave gap. The helical magnetic structure of FeP twists the spins within the $ab$ plane. Owing to the orthorhombic symmetry, the anisotropy matrix is allowed to have different components along the $a$ and $b$ axes within the $ab$ plane, which defines an anisotropy ellipse. The matrix $A$ then has the components:
\begin{equation}\label{eq5}
A =
\begin{pmatrix}
-D & 0 & 0\\
0 & 0 & 0\\
0 & 0 & D
\end{pmatrix}
,
\end{equation}
where $D$ is the anisotropy constant.

The ratio $J_{\text{c3}}/J_{\text{c0}}$ is fixed to reproduce the experimental $k$ of 0.2~r.l.u. According to Eq.~(\ref{eq:eq3}), this requires $J_{\text{c3}} \approx 0.618J_{\text{c0}}$. Although $J_{\text{b}}$ does not directly influence the simulated spectra for the momenta along $H$ and $L$, it becomes important for the resolution effects. To simulate the TAS spectra, we assumed that $J_{\text{b}} = 10J_{\text{c0}}$, which is justified based on the steep dispersion of Fig.~\ref{ris:fig2}(b). Furthermore, we found that it is necessary to include one additional interaction in the minimal model. The second-neighbor exchange along $a$, $J_{\text{a2}}$, is not needed to stabilize the magnetic structure of FeP but required to correctly reproduce the spectrum in the vicinity of the BZ boundary along $H$.

Figures~\ref{ris:fig5}(e)--\ref{ris:fig5}(h) show the side by side comparison of the simulated spectra to the TAS data of Figs.~\ref{ris:fig5}(a)--\ref{ris:fig5}(d). The calculated INS intensity (the magnon spectral weight) taking into account the resolution broadening in the energy and the two perpendicular momenta is shown as color maps. On top of the simulated INS intensity, the calculated magnon dispersions $\epsilon\left(\mathbf{q}\right)$ are shown as solid lines (only modes that have a significant spectral weight are shown for clarity).

First, one can notice that the relative intensities of the low-energy excitations at the spiral propagation vector are well captured in our simulations for all the compared reciprocal-lattice points.  While the INS intensity is altered between the ($11L$), $(10L)$, and $(00L)$ momenta, the calculated dispersions are identical, as expected for the equivalent momentum directions.

The spin-wave dispersions obtained for our model Hamiltonian in Eq.~\ref{eq4} within the linear spin-wave theory can be classified as follows. The magnetic helical structure of FeP yields three sets of modes. These are the dispersions $\epsilon^{\pm k,0}_{\mathbf{q}}$ that are the exact replica modes shifted along $L$ by $\pm k$ with respect to the zone center. This is depicted by different colors of the solid lines in Figs.~\ref{ris:fig5}(e)--\ref{ris:fig5}(g). Within each set, there are four modes: two ``acoustic'' modes labelled as $\text{A}_1$ and $\text{B}_1$, and two ``optic'' modes, which we denote as $\text{A}_2$ and $\text{B}_2$ (see Supplemental Materials~\cite{SM}). The modes $\text{A}_2$ and $\text{B}_2$ [omitted in Figs.~\ref{ris:fig5}(e)--\ref{ris:fig5}(g) because of their vanishing spectral weight] can be viewed as the replicas of the $\text{A}_1$ and $\text{B}_1$ dispersions shifted in momentum space along $L$ by $\pm 1$ if one considers the unfolded BZ~\cite{SM}. Thus, the solution of our Hamiltonian, in fact, consists of only two irreducible magnon modes.

The intensity distribution in our experimental spectra can be then readily interpreted. The low-energy part of the mode $\text{B}_1$ gains its highest spectral weight at the $q = 0$ replica at the (110) and (101) zones. Thus the $\cap$-shaped mode around the $\Gamma$ point is associated with the $\text{B}_1^{0}$ branch. The high-energy sector of the $\text{B}_1$ mode is predicted to exhibit high spectral weight at the $\text{B}_1^{+k}$ and  $\text{B}_1^{-k}$ replicas in the (002) zone [Fig.~\ref{ris:fig5}(g)]. However, because of kinematic constraints no experimental data are available for $E > 15$~meV at (002) for a direct comparison. In contrast to the $\text{B}_1$ mode, the mode $\text{A}_1$ becomes bright at the $q = \pm k$ replicas and has a vanishing intensity at the $q = 0$ replica. Thus, the $\text{A}_1^{\pm k}$ modes are responsible for the overall W-shape of the two crossing V-shaped modes seen in our INS data.

The simulated spectra for the $H$ direction are shown in Fig.~\ref{ris:fig5}(h). As was seen from the dispersions along $L$, three modes exhibit high spectral weight at the momentum that corresponds to the spiral propagation vector in the (101) zone. The modes $\text{B}_1^0$ and $\text{A}_1^{-k}$ have very similar upward dispersions along $H$ and reach the energy of $\sim$25 and 30~meV at the reduced momentum $q = 0.5$. The upper mode, $\text{A}_1^{+k}$, disperses upward from the energy of $\sim$27~meV at the zone center to $\sim$33~meV at $q = 0.5$. All the modes disperse downward for the reduced momenta $q > 0.5$ forming a local minimum at $q = 1$. The latter minimum is solely driven by the $J_{\text{a}2}$ exchange interaction in our model, which makes it important for a correct reproduction of the observed spectra along $H$.

\begin{table}[t]
\small\addtolength{\tabcolsep}{+7pt}
\caption{The free parameters and the parameters that were assumed as fixed of the model Hamiltonian in Eq.~(\ref{eq4}) used to reproduce the observed magnon spectra for the momenta along $H$ and $L$, in meV.}
\label{tab:tab1}
 \begin{tabular}{c c c c| c c}
 \midrule\midrule
  \multicolumn{4}{c}{free}  & \multicolumn{2}{c}{fixed} \\
\midrule
$J_{\text{a1}}S$ & $J_{\text{a2}}S$  & $J_{\text{c0}}S$ & $DS$ & $J_{\text{c3}}$ & $J_{\text{b}}$\\ [1ex] 
 \midrule
 2.25 & $-1.125$ & 11.25 &  0.05 & $0.618\:J_{\text{c0}}$ & $10\:J_{\text{c0}}$ \\ [1ex]   
 \midrule\midrule
\end{tabular}
\end{table}

\section{Discussion}\label{disc}

The magnon spectrum of FeP revealed in our INS measurements showed a very pronounced anisotropy of the spin-wave stiffness with respect to the main crystallographic axes. The magnon dispersion features a steep slope along the $K$ direction and is much softer and isotropic with respect to the other two reciprocal-space directions, $H$ and $L$. Thus, the surface of constant magnon energy at low energies takes a shape of a very oblate spheroid in three-dimensional momentum space, which points to an apparent quasi-one-dimensional character of the magnetic interactions in FeP. The observed excitations along $K$ in the vicinity of the zone center can be extrapolated to the entire BZ within our model. Such an extrapolation predicts the overall magnon bandwidth of $\sim$500~meV (see Supplemental Materials~\cite{SM}). 

On the one hand, the reduced dimensionality rarely emerges from the magnetic subsystem that is formed by a structurally three-dimentional network of magnetic ions, though such a phenomenon is observed here not for the first time~\cite{Nikitin21,Wu19}. On the other hand, the spin model proposed in the work of Kallel \textit{et al.}~\cite{Kallel74} indeed predicts a negligible exchange interaction for the ions coupled along the crystallographic $c$ axis ($J_{\text{a}} \gg J_{\text{c}}$). Because the spin-spiral ground state is independent from the spin-spin coupling along $b$ in this model, the model cannot predict if the magnetic interactions in FeP are one- or two-dimensional. The former would take place if the exchange along $b$ was comparably small to that along $c$ ($J_{\text{b}} \simeq J_{\text{c}}$), whereas the latter would correspond to the case $J_{\text{b}} \simeq J_{\text{a}}$. 

Surprisingly, the reduced dimensionality of magnetic interactions in FeP turns out to occur along a different route. As our experiments show, it is composed of FM spin chains along $b$ instead of the AFM spin chains along $a$ predicted by Ref.~\cite{Kallel74}. Consequently, the magnetic susceptibility of FeP should exhibit the behaviour typical for one-dimensional Heisenberg ferromagnets at temperatures comparable to the dominating exchange $J_{\text{b}}$. Our crude extrapolations made by the low-energy part of the magnon spectra along the $K$ direction yield $J_{\text{b}} \sim 112.5$~meV~$\approx 1300$~K, which is so high that it is close to the melting temperature of the compound.

The extracted free parameters of our model Hamiltonian~[Eq.~(\ref{eq4})] are listed in Table~\ref{tab:tab1}. The best agreement between the simulations and the data was achieved when the dominating exchange interaction in the $ac$ plane is an AFM coupling of the nearest-neighbor sites, $SJ_{\text{c0}} = 11.25$~meV. The nearest-neighbor exchange along $a$ is found to be five times smaller, $J_{\text{a1}} = 0.2\:J_{\text{c0}}$, whereas the next-nearest interaction along $a$, $J_{\text{a2}}$, is ferromagnetic and twice smaller than $J_{\text{a1}}$. The single-ion anisotropy constant $D$ is much smaller than the exchange energy and amounts to 0.05~meV$/S$, which is typical for $3d$ metals.

It is instructive to compare our results on FeP with the existing data available on its related compound FeAs. The latter crystallizes into the same space group with similar crystal-lattice parameters but a noticeably ($\sim$20\%) larger unit-cell volume due to the proportional expansion of all the lattice constants~\cite{Rodriquez11}. The double-helical structure of FeAs has a larger propagation vector of 0.395~r.l.u., also oriented along $c$. The angle between the two spirals $\alpha$ becomes larger in the case of FeAs and amounts to 26$^{\circ}$~\cite{Selte72}. The INS studies on FeAs~\cite{Niedziela21} showed that the spin excitations are nearly isotropic in the $ac$ plane with the bandwidth of $\sim$35~meV at low temperature. Because this closely resembles the results of our findings on FeP, one can expect that the spin model of FeP might be applicable also for FeAs. However, the lack of data on the magnon spectra along the $K$ direction in FeAs does not allow one to conclude if the spin excitations in FeAs are also low-dimensional, as in FeP, or in essence 3D in contrast to it.

Furthermore, it was shown in Ref.~\cite{Niedziela21} that FeAs exhibits soft excitations for momenta along $H$ at elevated temperatures just above $T_{\text{N}}$, such that the magnon spectra become one-dimensional with respect to the ($H0L$) plane in reciprocal space. This indirectly suggests that in FeAs the exchange interaction along $a$ is much weaker than that along $c$, which further indicates that FeAs and FeP are very similar materials with respect to their spin subsystem. The applicability of the spin model discussed in our work to the other members of the double-helix magnets, namely, MnP and CrAs, can be addressed in future studies.

\section{Conclusion}\label{concl}

To conclude, we conducted comprehensive inelastic neutron scattering measurements of the double-spiral helimagnet FeP in a broad range of momentum and energy transfer. The collected data revealed spin-wave excitations, which were resolved in the entire BZ along the main high-symmetry crystallographic directions. The typical magnon bandwidths are nearly isotropic in the reciprocal $(H0L)$ plane showing that the relevant exchange interactions between the neighboring spins along $a$ and $c$ are of the same order of magnitude. Surprisingly, the magnon dispersion along the $b$ axis demonstrates a much larger stiffness, making the whole spectrum strongly anisotropic. The apparent one-dimensional character of the magnon spectrum in FeP is rather surprising due to the apparent 3D crystal structure of the compound. Because the dominating exchange interaction along the $b$ axis is FM, the magnetic subsystem of FeP can be described as the FM spin chains with a weak AFM interchain coupling.

The key feature of the spin subsystem of FeP is the magnetic frustration that leads to stabilization of the spin-spiral order. The obtained spectra allowed us to closely examine the applicability of the spin interaction model that was proposed in the previous studies~\cite{Kallel74}. While this model is able to correctly predict the ground state, it fails to reproduce the observed excitations. In order to describe the experimental spectra, we proposed an effective model that is based on a different frustration mechanism. The spin-wave simulation within the new model showed excellent agreement with the INS data, enabling us to quantify the most important exchange interactions that couple the spins along $a$ and $c$. Our model can be used in future studies of the similar double-helix magnets, such as FeAs, CrAs, and MnP.

\section*{Acknowledgements}

We thank I. Mazin for fruitful discussions. We acknowledge support from the German Research Foundation (DFG) under Grants No. SA 523/4-1 and IN 209/9-1, via the projects C03 and C04 of the Collaborative Research Center SFB 1143 (project-id 247310070) at the TU Dresden and the W\"{u}rzburg-Dresden Cluster of Excellence on Complexity and Topology in Quantum Materials\,---\,\textit{ct.qmat} (EXC 2147, project-id 390858490). S.E.N. acknowledges funding from the European Unions Horizon 2020 research and innovation program under the Marie Sk\l{}odowska-Curie grant agreement No. 884104.

\end{document}